# Benchmark 3-Flavor Pattern and Small Universal Flavor-Electroweak Parameter

E. M. Lipmanov

40 Wallingford Road # 272, Brighton MA 02135, USA

**Abstract**

The electroweak theory contains too many empirical parameters. Most of them are related to the flavor part of particle physics. In this paper we discuss a relevant simple idea: the complicated system of actual dimensionless, small versus large, quantities in elementary particle flavor phenomenology is considered as small deviated from an explicitly defined 'benchmark' flavor pattern with no tuning parameters. One small empirical universal dimensionless parameter measures this deviation. Its possible physical connections are discussed. As inferences, quasi-degenerate neutrino type with mass scale m $\cong$ 0.16 –0.18 eV, neutrino and quark mixing matrices, large neutrino oscillation 3-flavor hierarchy, r $\cong$ 0.034, and quark-neutrino complementarity are predicted.

## 1. Introduction

The highly successful one-generation (e.g. electron generation) electroweak theory (EWT) [1] establishes two small electric e and semi-weak $g_W$ charges imposed on a SU(2)-dublet of electron and neutrino with masses $m_e$ and $m_\nu$. Gauged symmetry does not determine the magnitudes of the charges and there are no relations between primary (bare) particle masses and charges. The one-generation EWT extended to three generations [2], [3] and [4] contains more than threefold enlarged number of parameters in the empirical particle mass matrices. To reduce the number of free parameters in EWT is one of the most needed tasks.

In Sec.2, three particle generations are considered as necessary for a substantial connection between dimensionless flavor quantities and electroweak charges.

In Sec3, the benchmark flavor pattern is defined.

In Sec.4, realistic particle flavor pattern is discussed.

In Sec.5, physical meaning of the one small universal $\varepsilon$-parameter is discussed. Sec.6 contains conclusions.

## **2. 3-flavor DMD-hierarchies and electroweak charges**

The semi-empirically extended by flavor EWT [2], [3] and [4] describes particle 3-flavor connections by mixing matrices (quark and neutrino ones) that are not related to the basic dimensionless quantities - electroweak charges.

An apparent radical way to substantially reduce the number of dimensionless parameters in the EWT is to connect them if possible with the electroweak charges. This is indeed possible in low energy phenomenology by taking into account the empirical fact of additional degree of particle freedom generated by particle mass copies (flavor generations). Dimensionless-made particle electroweak charges may be connected to mass ratios of these copies, or rather to the richer in physical meaning mass-degeneracy deviation (DMD) quantities and especially to hierarchies of these DMD-quantities [5]. Then, particle charges, though not related to individual masses, may be connected with the particle-copy mass ratios - it means that flavor particle generations are needed to enhance the unity of EW physics by *connection of EW charges with particle-copy mass distributions*.

Since there are only two independent low energy electroweak charges, unique connection of the particle DMD-hierarchies with electroweak charges requires three low energy particle flavor copies i.e. the fact of two electroweak charges points to three particle flavor generations so that the relation between one-generation EWT and flavor physics may be unique and on an essential physical level. In terms of mass matrix elements it means that particle mass ratios and mixing angles have to be expressed through the electroweak charges.

As discussed below, the idea of unique relations between the two (charge lepton and neutrino) 3-flavor DMD-hierarchies and two electroweak charges can be remarkably, though approximately, realized in a phenomenological semi-empirical flavor pattern with one new small dimensionless parameter $\varepsilon$ related to the electron charge. That parameter $\varepsilon$ measures the deviations of the actual flavor quantities from the ones of a defined 'benchmark' flavor pattern serving as a necessary background.

### **3. Benchmark flavor pattern of elementary particles**

An ingenuous view of the elementary particle flavor pattern from known experimental data of mass ratios and mixing angles[1] can be described by the following benchmark flavor pattern, with quark (q) and neutrino ($\nu$) mixing matrices,

$$[m_e \neq 0,\ m_\nu \cong 0,\ m_\mu,\ m_\tau,\ m_q \cong \infty],$$

$$m_\nu/m_e \cong 0,\ m_e/(m_\mu,\ m_\tau,\ m_q) \cong 0, \qquad (1)$$

$$\begin{pmatrix} 1 & 0 & 0 \\ 0 & 1 & 0 \\ 0 & 0 & 1 \end{pmatrix}_q, \quad \begin{pmatrix} 1/\sqrt{2} & 1/\sqrt{2} & 0 \\ -1/2 & 1/2 & 1/\sqrt{2} \\ 1/2 & -1/2 & 1/\sqrt{2} \end{pmatrix}_\nu. \qquad (2)$$

It describes an extreme presentation of the main feature of 'small' versus 'large' quantities of experimental flavor data: zero quark mixing, maximal neutrino mixing, zero 'reactor' neutrino angle, degenerate massless neutrinos and very large Dirac particle masses above the electron mass.

---

[1] CP-violating phases not considered here.

Advantage of this extreme pattern is being the benchmark for the realistic flavor pattern. At leading approximation, it appears enough to have one small dimensionless $\varepsilon$-parameter: small flavor quantities are described by powers $\varepsilon^n$ while large ones - by powers $(1/\varepsilon)^n$ with n an integer.

From comparison with experimental data it follows that the benchmark flavor pattern (1)-(2) can be represented as zero approximation of expansion of the realistic flavor quantities in terms of the small parameter $\varepsilon$, which magnitude is approximately given by

$$\varepsilon \cong 0.082 \cong \exp(-5/2). \qquad (3)$$

## **4. Realistic elementary particle flavor pattern**

Small $\varepsilon$-deviations of the realistic particle flavor pattern from the benchmark one (at tree EW approximation) should maintain its main overall 'small versus large' flavor features leading to finite but small quark mixing, small deviation from maximal neutrino mixing, quasi-degenerate (QD) neutrinos, finite extraordinary small neutrino masses, finite but large (in comparison with electron mass) charged lepton (CL) and quark masses and approximate quark-neutrino mixing complementarity [6].

The CL mass ratios and quark and neutrino mixing angles are expressed through the $\varepsilon$-parameter in simple form [5]:

$$\cos^2 2\theta_{12} \cong \sin^2 2\theta_c \cong (2\sqrt{2})(m_\mu/m_\tau) \cong 2\varepsilon, \qquad (4)$$

$$\cos^2 2\theta_{23} \cong \sin^2 2\theta' \cong (2\sqrt{2})(m_e/m_\mu) \cong 2\varepsilon^2. \qquad (5)$$

Here $\theta_{12}$ and $\theta_{23}$ are the solar and atmospheric neutrino oscillation large mixing angles, $\theta_c$ and $\theta'$ are the quark Cabibbo angle and next to it mixing one.

It should be noted that relations (4) and (5) originally resulted from two semi-empirical flavor 'rules' - 1) universal quadratic DMD-hierarchy rule between generic flavor pairs (1 and 2) of DMD-quantities, namely $[DMD\,2]^2 \cong 2[DMD\,1]$, which quantitatively connects all relations (4) with the respective relations (5) see [5], and 2) Dirac-Majorana DMD-duality rule for Majorana neutrinos; it relates e.g. large CL DMD-quantities

$$[DMD\,2] = [(m_\tau^2/m_\mu^2) - 1] \cong 2/\varepsilon^2, \quad [DMD\,1] = [(m_\mu^2/m_e^2) - 1] \cong 2/\varepsilon^4 \gg 1, \tag{6}$$

to small neutrino DMD-quantities

$$[DMD\,2] = [(m_3^2/m_2^2) - 1] \cong 2r, \quad [DMD\,1] = [(m_2^2/m_1^2) - 1] \cong 2r^2 \ll 1, \tag{7}$$

where $m_1 < m_2 < m_3$ are organized neutrino masses and $r$ is the neutrino oscillation (solar-atmospheric) hierarchy parameter approximately expressed in terms of the $\varepsilon$-parameter as given by

$$r = \Delta m^2_{sol}/\Delta m^2_{atm} \cong -\varepsilon^2 \log \varepsilon^2 \cong 1/30. \tag{8}$$

The phenomenon of quark-neutrino complementarity [6],

$$2\theta_{12} \cong (\pi/2 - 2\theta_c), \quad 2\theta_{23} \cong (\pi/2 - 2\theta'), \tag{9}$$

is described by the relations (4) and (5) and has here a simple physical meaning. It comes from the idea that realistic mixing of elementary particles appears as a small deviation from the benchmark flavor mixing pattern, which is determined by only one small parameter $\varepsilon$.

From (4) and (5), the quark mixing-matrix elements are

$$c_{12} \cong (1 - \varepsilon/4), \quad s_{12} \cong \sqrt{(\varepsilon/2)}, \quad c_{23} \cong (1 - \varepsilon^2/4),$$

$$S_{23} \cong \varepsilon/\sqrt{2}, \quad S_{13} \cong \varepsilon^2/2 . \tag{10}$$

The notations here are $C_{12} = \cos\theta_c$, $S_{12} = \sin\theta_c$, $C_{23} = \cos\theta'$, $S_{23} = \sin\theta'$, $S_{13}$ is the amplitude of the CP-violating term in the CKM mixing matrix[2] [4].

With (10) the realistic quark mixing matrix is approximately given by

$$V_q \cong \begin{pmatrix} (1-\varepsilon/4) & \sqrt{(\varepsilon/2)} & \varepsilon^2/2 \\ -\sqrt{(\varepsilon/2)} & (1-\varepsilon/4) & \varepsilon/\sqrt{2} \\ \varepsilon\sqrt{\varepsilon}/2 & -\varepsilon/\sqrt{2} & 1 \end{pmatrix}_q \cong \begin{pmatrix} 0.98 & 0.2 & 0.0034 \\ -0.2 & 0.98 & 0.058 \\ 0.0003 & -0.058 & 1 \end{pmatrix}, \tag{11}$$

in fair agreement with the CKM data values [4].

From relations (4) and (5) follow also the magnitudes of neutrino mixing matrix elements

$$C_{12} \cong \sqrt{[(1+\sqrt{(2\varepsilon)})/2]}, \quad S_{12} \cong \sqrt{[(1-\sqrt{(2\varepsilon)})/2]},$$

$$C_{23} \cong \sqrt{[(1+\varepsilon\sqrt{2})/2]}, \quad S_{23} \cong \sqrt{[(1-\varepsilon\sqrt{2})/2]}, \quad S_{13} \cong \varepsilon^2/2. \tag{12}$$

The notations are $C_{12} = \cos\theta_{12}$, $S_{12} = \sin\theta_{12}$, $C_{23} = \cos\theta_{23}$, $S_{23} = \sin\theta_{23}$, $S_{13} = \sin\theta_{13}$ is supposed equal to the corresponding quark matrix element.

From (12), the realistic neutrino mixing matrix is approximately given, in the standard representation [4], by

$$V_\ell \cong \begin{pmatrix} 0.84 & 0.55 & 0.0034 \\ -0.41 & 0.62 & 0.66 \\ 0.36 & -0.56 & 0.75 \end{pmatrix}_\nu . \tag{13}$$

It is close, but not equal to the Harrison-Perkins-Scott (HPS) 'tribimaximal' matrix [7]; the deviation of the atmospheric neutrino oscillation parameter $S_{23}$ from the maximal HPS-value is finite though small ~6%; the deviation of the solar neutrino oscillation parameter $S_{12}$ from maximal mixing is ~20% while the deviation from the HPS-value is small ~4%.

---

[2] Notations '$C_{ij}$' and '$S_{ij}$' for matrix elements are as in the CKM-matrix representation [4].

Estimation of absolute QD-neutrino mass scale follows from the relations (7) for neutrino DMD-quantities, e.g.

$$m_\nu \cong (\Delta m^2_{sol}/2r^2)^{1/2}. \qquad (14)$$

With best fit values from the data analysis [12] of solar mass-squared difference and hierarchy parameter $r$, $(\Delta m^2_{sol})_{bf} \cong 7.6 \times 10^{-5}$ eV$^2$, $r_{bf} = 0.032$, and also $3\sigma$ ranges $\Delta m^2_{sol} \cong (7.1 - 8.3) \times 10^{-5}$ eV$^2$, $r \cong (0.027 - 0.040)$, quantitative estimations for absolute QD-neutrino mass scale are

$$(m_\nu)_{bf} \cong 0.18 \text{ eV}, \quad (m_\nu)_{3\sigma} \cong (0.15 - 0.24) \text{ eV}. \qquad (15)$$

It should be noted an interesting feature of the benchmark flavor pattern revealed by the fact that available data for known dimensionless flavor quantities are quantitatively described by small integer powers of the universal parameter $\varepsilon^n$, $n = 1 \div 6$, with coefficients close to common numbers [8] '2' and 'π'. This statement is exemplified above by CL and neutrino mass ratios and quark and neutrino mixing matrix elements and in examples below.

Let us consider some other flavor quantities.

1) QD-neutrino-electron mass ratio [9]. At the benchmark (1) (at $\varepsilon = 0$) this mass-ratio is zero; so at $\varepsilon \neq 0$ it should be

$$m_\nu/m_e = a\,\varepsilon^x, \quad x > 0, \qquad (16)$$

where $(a, x)$ are two unknowns and $m_\nu$ is the average QD-neutrino mass $m_\nu = (m_1 + m_2 + m_3)/3 \equiv \Sigma m_\nu/3$. By comparison (16) with the semi-empirical estimation of neutrino mass from oscillation data in (15) it follows $x = 6$ and $a = (0.96 \div 1.53)$ from $3\sigma$-ranges, and $a = 1.15$ from the best fit

estimation (15). It suggest the value a = π/3 with result given by

$$3m_v/m_e \cong \pi\varepsilon^6, \quad m_v \cong 0.16 \text{ eV}. \qquad (17)$$

Coefficient '3' on the left has physical meaning of three QD-neutrino masses, so the left side of first relation (17) means ratio of the QD-neutrino mass sum to electron mass.

The sum of three QD-neutrino masses is $\Sigma m_v = (0.50 \pm 0.003)$ eV; it fits the astrophysical constraints [10]

$$(\Sigma m_v)_{exp} < 0.61 \text{ eV} \quad (95\% \text{ CL}). \qquad (18)$$

2) Electron-top-quark mass ratio [9]. At $\varepsilon = 0$ (1) this mass-ratio is zero; at $\varepsilon \neq 0$ it should be

$$m_t/m_e = (b\,\varepsilon^y), \quad y < 0, \qquad (19)$$

where (b, y) are two unknowns. By comparison with experimental data on t-quark mass [4] we get

$$3m_t/m_e \cong 1/\pi\varepsilon^6, \quad m_t \cong 177.2 \text{ GeV}. \qquad (20)$$

The first relation in (20) has notable physical meaning – ratio of the sum of three color degenerate top-quark masses to electron mass. The magnitude of top-quark mass $m_t$ in (20) agrees with data values [4] to within ~1 S.D.

A basic 'geometric' relation between the sums of three degenerate QD-neutrino and top-quark masses follows from the semi-empirical estimations (17) and (20), generated by the idea of benchmark flavor pattern (1)-(2),

$$(3\,m_v)(3\,m_t) = m_e^2. \qquad (21)$$

It is a semi-empirical geometric seesaw-like connection between neutrino and top-quark masses with electron mass $m_e$ at the geometric middle of the two extreme elementary particle masses[3]. Unlike (17) and (20), this interesting relation [9] does not contain empirical parameters or coefficients. Note that the equal coefficients 3 on the left side of (21) in the neutrino and t-quark terms have different physical meaning – the t-quark coefficient means three colors, but the neutrino one means three flavors.

---

[3] It seems as if it closed the low energy island of elementary particles.

Top-quark mass $m_t$ in (21) is the large counterpart (close to the electroweak scale) of the small neutrino mass. Such representation is possible only for QD-neutrinos and equal numbers of particle flavors and quark colors.

Small neutrino mass versus large top-mass and reversely analogous phenomenon of large neutrino mixing versus small quark one are two distinct features of the way QD-neutrinos fit the pattern of elementary particle masses[4].

The absolute QD-neutrino mass from the top-quark data is given by

$$m_\nu = m_e^2/3\,m_t \cong 10^{-12}\,m_t \cong 0.17\,\text{eV}. \qquad (22)$$

3) $\varepsilon$-hierarchical pattern of heavy quark masses. The considered above semi-empirical approach to electron-Dirac-particle (CL and t-quark) mass ratios as small deviations (small $\varepsilon$–parameter) from their benchmark values, at $\varepsilon = 0$, leads to the following estimations of three heavy quark masses:

$$m_e/3m_s \cong \pi\varepsilon^3, \quad m_s \cong 98 \text{ MeV}. \qquad (23)$$

$$m_e/3m_c \cong \pi\varepsilon^4, \quad m_c \cong 1.19 \text{ GeV}, \qquad (24)$$

$$m_e/3m_b \cong \pi^2\varepsilon^5, \quad m_b \cong 4.6 \text{ GeV}, \qquad (25)$$

in fair agreement with PDG data [4]. All these are relations between *sums* of degenerate quark and lepton masses; the factor '3' appears explicitly only in quark-lepton or neutrino-CL relations, not in quark, CL or neutrino mass rations themselves.

---

[4] For a connection between small quark mixing, large neutrino mixing and QD-neutrino type in a gauge model see [11].

And so, the semi-empirical mass ratios of electron mass $m_e$ to sums of color degenerate heavy quark masses $3m_k$ are approximately described by the ε-hierarchical pattern

$$m_e/3m_k \cong \pi\varepsilon^k, \quad m_e/3m_b \cong \pi^2\varepsilon^5, \qquad (26)$$

for the strange, charm and t-quark k = 3, 4, 6 respectively with an exception for the bottom quark where the ε-power value is regular k = 5, but the coefficient is $\sim\pi^2$ (or $3\pi$) instead of regular $\pi$.

## 5. On physical meaning of the universal flavor ε–parameter

**1.** DMD-quantities and hierarchies of CL and neutrinos are basic dimensionless observable quantities in lepton flavor physics. There are two DMD-quantities and one DMD-hierarchy for three charged leptons and same for the three neutrinos:

$$\text{DMD(CL)}1 = [(m_\tau^2/m_\mu^2)-1], \quad \text{DMD(CL)}2 = [(m_\mu^2/m_e^2)-1], \qquad (27)$$

$$\text{DMDH(CL)} = \text{DMD(CL)}1 / \text{DMD(CL)}2; \qquad (28)$$

$$\text{DMD}(\nu)1 = [(m_2^2/m_1^2)-1], \quad \text{DMD}(\nu)2 = [(m_3^2/m_2^2)-1], \qquad (29)$$

$$\text{DMDH}(\nu) = \text{DMD}(\nu)1 / \text{DMD}(\nu)2. \qquad (30)$$

By definition, the magnitude of CL mass-squared DMD-hierarchy DMDH(CL) from (27)-(28) and experimental data for CL masses and fine structure constant α [4] –

$$\text{DMDH(CL)} \cong \varepsilon^2 \cong \alpha \qquad (31)$$

is close (~10%) to the magnitude of α at pole value of the photon propagator $q^2 = 0$.

The DMD-hierarchy quantity of QD-neutrinos from (29)-(30) and (8) is given by

$$\text{DMDH}(\nu) \cong r \cong 5\varepsilon^2 \cong 0.0337, \qquad (32)$$

it is equal to the solar-atmospheric hierarchy parameter $r$ from neutrino oscillation data and its magnitude is close to $5\varepsilon^2$. On the other hand, the quantity $5\varepsilon^2$ is close to the semi-weak analog $\alpha_W = g_W^2/4\pi$ of the fine structure constant $\alpha$ at pole value of the Z-boson propagator from PDG data [4]:

$$\alpha(M_Z) = 1/(128.91 \pm 0.02), \quad (\sin^2\theta_W)|_{M_Z} = 0.23108 \pm 0.00005,$$
$$\alpha_W(M_Z) = (\alpha/\sin^2\theta_W)|_{M_Z} \cong 0.0336. \qquad (33)$$

$\theta_W$ is the Weinberg mixing angle. The numbers in (32) and (33) in the relation $\alpha_W(M_Z) \cong 5\varepsilon^2$ disagree only by ~0.3%.

The best-fit values and $3\sigma$ allowed ranges for solar-atmospheric hierarchy parameter from the three-flavor neutrino oscillation global data analysis are given by [12]

$$r_{bf} = 0.032, \quad 0.027 \leq r_{3\sigma} \leq 0.040. \qquad (34)$$

The suggested quantitative connection between oscillation hierarchy parameter $r$ and flavor parameter $\varepsilon$

$$r \cong -\varepsilon^2 \log \varepsilon^2 \cong 0.0337 \qquad (35)$$

is well within the $3\sigma$ ranges (34) and agrees with the best fit value $r_{bf}$ from (34) to within an accuracy ~5%, compare (31).

If further confirmed by experimental data with higher confidence for both the $r$-parameter at neutrino oscillation experiments and the electroweak interaction constant $\alpha_W(q^2 = M_Z^2)$ at Z-pole experiments, the relation between neutrino DMD-hierarchy, oscillation hierarchy parameter and the semi-weak coupling constant squared,

$$DMDH(\nu) \cong r \cong \alpha_W(M_Z) \cong 0.034, \qquad (36)$$

will be a suggestive evidence in favor of QD-neutrino type as resulted from new fundamental physics.

Relation (36) may be the true physical meaning of the neutrino oscillation solar-atmospheric hierarchy parameter *r*. This relation is a quantitative answer of why is the solar neutrino mass-squared difference much smaller than the atmospheric one. In that regard it is relevant to address the case of not-QD-neutrinos. In general with the choice $m_1 < m_2 < m_3$ it follows

$$DMDH(\nu) \cong (m_1^2/m_2^2)\, r. \qquad (37)$$

For the physical meaning of solar-atmospheric hierarchy parameter *r* to be the unique DMD-hierarchy DMDH($\nu$), the condition

$$(m_1^2/m_2^2) \cong 1 \qquad (38)$$

must be fulfilled.

In case of inverse neutrino mass ordering ('hierarchy') the condition (38) means back to QD-neutrinos. But there is a special case of not-QD-neutrinos - with 'normal' neutrino mass ordering and relations[5]

$$\Delta m^2_{sol} << m_1^2,\, m_2^2 < \approx \Delta m^2_{atm}, \qquad (39)$$

where the condition DMDH($\nu$) $\cong r$ may be also approximately fulfilled. In this particular not-QD-neutrino case the solar-atmospheric hierarchy parameter *r* has the more general[6] physical meaning of hierarchy of deviations from mass-degeneracy between the two pairs of neutrino masses ($m_2$, $m_1$) and ($m_2$, $m_3$),

$$DMDH(\nu_{N-QD})\big|_{(38)} \cong r, \qquad (40)$$

as it is in the QD-case. But in not-QD case the relation (40) seems an accidental one (normal ordering and restricted neutrino mass interval as tuning conditions) in contrast to QD-neutrino case where it is fulfilled by the QD-definition. In any case, the importance of further verification of the new physics suggestive relation (36) between the neutrino DMD-hierarchy, solar-atmospheric parameter $r$ and electroweak constant $\alpha_W(M_Z)$ cannot be overestimated.

---

[5] For example $m_2^2 \cong m_1^2 \cong (0.2 \div 4) \times 10^{-3}$ eV$^2$; $\Delta m^2_{atm} \cong 2.4 \times 10^{-3}$ eV$^2$ [12].

[6] Unlike the primary definition of the solar-atmospheric hierarchy parameter *r*, DMD-hierarchy is a general physical notion applicable to 3-flavor mass systems such as leptons and quarks.

The suggested understanding of QD-neutrino oscillation hierarchy-parameter $r$ as DMD-hierarchy quantity (32) is independent of the oscillation particulars; $r \cong \text{DMDH}(\nu)$, points to possible relation to the electroweak interaction constant, namely $r \cong \alpha_W(q^2 = M_Z^2)$, in analogy with the CL relation (31) $\text{DMDH(CL)} \cong \alpha$.

**2.** The constant $\varepsilon^2$ is close (~92%) to the fine structure constant $\alpha$ at zero momentum transfer; but more, there are highly accurate factual empirical relations [13] between the two constants $\varepsilon$ and $\alpha(q^2 = 0)$:

i) $\qquad\qquad 1/\varepsilon^2 \cong (\exp\alpha\,/\alpha)^{\exp 2\alpha}, \qquad\qquad (41)$

this relation determines the parameter $\varepsilon = \exp(-5/2)$ to within $10^{-5}$ through the experimental value $\alpha_{\text{Data}}$ [4] of the fine structure constant[7];

ii) $\qquad (\exp\alpha\,/\alpha)^{\exp 2\alpha} + (\alpha/\pi) = 1/\varepsilon^2, \qquad (42)$

with $\varepsilon = \exp(-5/2)$ this relation is true to within $(\alpha - \alpha_{\text{Data}})/\alpha_{\text{Data}} \cong 10^{-8}$;

iii) $\qquad (\exp\alpha\,/\alpha)^{\exp 2\alpha} + [(\alpha/\pi) + O(\alpha^2)] = 1/\varepsilon^2 \qquad (43)$

with $\varepsilon = \exp(-5/2)$ and $O(\alpha^2) = -\alpha^2/4\pi + O(\alpha^3)$, this equation determines a solution for the fine structure constant $\alpha$

---

[7] With $\varepsilon = \exp(-5/2)$ the connections (41), (42) and (43) are accurate quantitative empirical statements. Without established successful flavor theory (like the one-generation standard model) it seems premature to consider them accidental in view of the discussion above.

with accuracy $10^{-10}$, see [13] and compare with new data analysis in ref. [14].

The above discussion suggests that particle 3-flavor physics and the highly successful one-generation electroweak theory are *essentially* connected through the universal small parameter ε: if ε → 0, the considered ratios of second and third generation particle masses to the electron mass would increase infinitely, so the extra particle generations would get unobservable. But at the same time the electroweak interactions and atomic bound states of the first generation particles would vanish: α → $ε^2$ → 0.

## 6. Conclusions

The subject of this paper is phenomenology of dimensionless quantities in flavor and electroweak physics that include bare particle mass ratios and mixing angles, and electroweak charges. The system of actual flavor quantities is considered against the background of a benchmark flavor pattern such that the realistic flavor pattern appears small deviated from the benchmark one. This deviation is approximately described by one small empirically emerging universal parameter, which is related in simple form to the dimensionless-made universal electric charge of the electron. The need of particle flavor degree of freedom with three generations, QD-neutrino type, absolute neutrino mass scale, neutrino and quark mixing matrices, physical meaning and magnitude of the neutrino oscillation hierarchy parameter and the phenomenon of quark-neutrino mixing complementarity are among considered main physical problems.